\documentclass[pra,preprint,tightenlines,showpacs,twocolumn,10pt]{revtex4} 
\usepackage{epsfig} 

\begin{document}  
    
\title{ Young-type interference in projectile-electron 
loss in energetic ion-molecule collisions  }
 
\author{ A.B.Voitkiv, B.Najjari and D.Fischer }   
\affiliation{ Max-Planck-Institut f\"ur Kernphysik, 
Saupfercheckweg 1, D-69117 Heidelberg, Germany } 
\author{ A.Artemyev and A.Surzhykov } 
\affiliation{ Physikalisches Institut, Universit\"at 
Heidelberg, Philosophenweg 12, D-69120 Heidelberg, Germany \\ 
and GSI Helmholtzzentrum f\"ur Schwerionenforschung 
GmbH, Planckstrasse 1, D--64291 Darmstadt, Germany} 


\begin{abstract}  

Under certain conditions an electron bound in 
a fast projectile-ion, colliding with a molecule, 
interacts mainly with the nuclei and inner shell electrons 
of atoms forming the molecule. 
Due to their compact localization in space and 
distinct separation from each other 
these molecular centers play in such collisions 
a role similar to that of optical slits in light scattering 
leading to pronounced 
interference in the spectra of the electron 
emitted from the projectile.    

\end{abstract} 


\maketitle 



The wave--particle duality, which states that all 
atomic objects exhibit particle as well as wave properties, 
is one of the basic concepts of quantum mechanics. 
Proposed initially by Louis de Broglie \cite{dBr92} in 1923, 
this concept has been confirmed few years later 
in the electron diffraction experiments \cite{DaG27,Tho28}. 
Since then, a large number of investigations have been performed 
in order to observe the wave nature of not only electrons 
but also heavier particles such as, for example, neutrons, atoms, dimers and 
even fullerenes $C_{60}$ \cite{Arn99}. Most of  these measurements 
were aimed at a demonstration of Young's double--slit phenomena, 
in which the coherent addition of the amplitudes of two (or many) 
paths, leading to interference, is related to the wave--like particle behavior.

In the atomic world the natural analog of the 
Young's slits is represented by diatomic molecules. 
Starting with the works \cite{Tuan}-\cite{CoF66}, 
especially significant interest has been focused on 
studying interference phenomena involving 
homo-nuclear molecules \cite{photo-laser}-\cite{double-capt-we}.  

These studies were dealing with two principally different 
interference scenarios. In one of them the attention was focused 
on interference in the spectra of electrons emitted from the molecule 
in the course of photoionization \cite{CoF66} - \cite{lin} 
and consequent Auger decay \cite{cherepkov},  
as well as in ionization by electrons \cite{e-2e} 
and heavy ions \cite{Sto01}-\cite{cristina}.  
Note that in such a scenario, unlike the Young's experiment, 
the wave is not diffracted by the "slits" but rather emerges from them.   
In the second scenario, which was realized in 
\cite{daniel}-\cite{double-capt-we} for electron capture 
and proton scattering  
and is a more direct analog of the Young's optical experiment, 
interference is caused by coherent scattering 
of the incident projectile on the atomic centers  
of the molecule.    

In this letter we propose yet another way to 
collision-induced interference. 
It falls into the second scenario but, similarly to 
\cite{Sto01}-\cite{cristina},  
deals with interference in electron emission spectra. 
It is realized in collisions of molecules with partially 
stripped multiply-charged projectile-ions, 
in which the electron(s) of the projectile is emitted. 

Compared to the electron emission, studied in \cite{Sto01}-\cite{cristina}, 
the present case possesses important differences.  
In particular, in the situation, considered in \cite{Sto01}-\cite{cristina},   
the electron wave is launched from the "slits", 
which are not really separated and well localized    
since the electrons of molecules like H$_2$ 
occupy the whole volume of the molecule and 
are mainly located not on the atomic nuclei 
but rather between them. As a result, 
the corresponding interference pattern is rather smooth. 
In contrast, as will be shown below, 
the emission from the projectile occurs due 
to a coherent scattering of the electron of the projectile 
on the {\it nuclei} of the molecule (partially screened by 
the inner shell electrons) and, therefore, 
the "slits" are very well separated and localized 
in space that can lead to very pronounced 
interference effects in the emission pattern.  


Below, based on the relativistic time--dependent 
perturbation approach,  
we shall derive the cross section for electron loss 
in collisions with homo-nuclear dimers.  
The possibility of interference effects will be demonstrated 
by calculating the cross section for   
fast hydrogen--like magnesium Mg$^{11+}$(1s) and 
S$^{15+}$(1s) ions 
colliding with N$_2$ dimers. 

Atomic units are used throughout except where 
otherwise stated. 

Since the collision between an ion carrying an electron 
and a molecule in general represents a very 
complex many-body problem, our consideration 
will be based on a simplified model which, however, 
takes into account all essential physics  
of the collision process in question. 
Within this model, in order to describe electron transitions 
in the projectile we shall use the first order 
perturbation theory in the interaction between 
this electron and the molecule. Such an approximation 
is a good one, provided $Z_p \stackrel{>}{\sim} Z_A $, 
where $Z_p$ and $Z_A$ are, respectively, 
the nuclear charges of the ion and the atoms in the molecule, 
and one merely wishes to describe projectile-electron transitions,  
without paying attention to what happens with 
the molecule in such collisions. 

Further, we shall only consider molecules whose atoms 
have relatively large atomic numbers, $Z_A \gg 1$. 
Under the simultaneously fulfilled conditions 
$Z_A \gg 1$ and $Z_p \stackrel{>}{\sim} Z_A $ 
the main contribution to the projectile-electron transitions  
in collisions with the molecule is given by the screening 
target mode, in which the projectile electron interacts with 
the molecule "frozen" during the short collision time 
in its initial state \cite{my-book}.  

Moreover, provided the condition $Z_p \stackrel{>}{\sim} v$ 
is fulfilled ($v$ is the collision velocity) the momentum 
transferred in the collision becomes so large 
(on the molecular scale) that the outer electrons 
of the molecule are not able to screen 
the nuclei of the molecule. Therefore, 
the main contribution to the electron loss 
arises from the interaction 
with the nuclei of the molecule partially screened 
by the inner shell electrons. 
Thus, the projectile electron undergoes transitions 
due to the interaction with well 
localized centers of force which, in addition, 
are well separated in space. 
Besides, since the inner electrons are basically 
atomic electrons, one can treat the molecule as 
a sum of free atoms and use the atomic parameters 
for the description of the field produced 
by the molecule in the collision. 

Taking all this into account, 
the scalar potential describing the field 
of the molecule in its rest frame $K'$ can be written as 
\begin{eqnarray} 
\Phi'_M({\bf r}') = \sum_{j=1}^2 \frac{Z_j \phi_j(|{\bf r}' - {\bf R}'_j|)}{|{\bf r}' - {\bf R}'_j|}, 
\label{e1}
\end{eqnarray}
where ${\bf r}'$ is the observation point of the field 
and ${\bf R}'_j$ is the coordinate 
of the nucleus of the $j$-th atom ($j=1,2$),  
$Z_j$ the charge of the nucleus and   
\begin{eqnarray} 
\phi_j(x) = \sum_{l} A_j^l \exp(-\kappa_j^l x) 
\label{e2}
\end{eqnarray} 
with the screening parameters  
$A_j^l$ ($\sum_l A_j^l = 1$) and $\kappa_j^l$ 
tabulated in \cite{Moliere} and \cite{Salvat}. 

It is convenient to treat the projectile-electron 
transitions using the reference frame $K$ 
in which the nucleus of the projectile is at rest. 
We take the position of the nucleus as the origin of $K$ 
and assume that in this frame the center of mass 
of the molecule moves along a straight-line 
classical trajectory ${\bf R}(t)={\bf b} + {\bf v} t$, 
where ${\bf b}=(b_x,b_y,0)$ is 
the impact parameter, ${\bf v}=(0,0,v)$ 
is the collision velocity and $t$ is the time.   
Using Eqs.(\ref{e1})-(\ref{e2}) and 
the Lorentz transformation for the potentials 
we obtain that the electromagnetic field 
of the molecule in the frame $K$ 
is described by the potentials 
\begin{eqnarray}
\Phi_M({\bf r},t) &=& %
\gamma \Phi'_M({\bf s}_j) 
\nonumber \\ 
{\bf A}_M({\bf r},t) &=& \left(0,0, \frac{v}{c} \Phi_M \right),  
\label{e4} 
\end{eqnarray}
where ${\bf r} = ({\bf r}_{\perp}, z)$ 
with ${\bf r}_{\perp} \cdot {\bf v} =0$ 
is the coordinate of the point of observation of the field 
in the frame $K$, $c$ is the speed of light 
and $\gamma=1/\sqrt{1-v^2/c^2}$ is the collisional Lorentz factor. 
Further,  
\begin{eqnarray} 
{\bf s}_j = \left( \gamma (z - v t_j), {\bf r}_{\perp} - {\bf b}_j \right),   
\label{e5} 
\end{eqnarray} 
where ${\bf b}_j={\bf b} + \delta {\bf b}_j $ 
is the impact parameter for the nucleus of the $j$-th atom 
of the molecule, $t_j$ is the time of its closest approach 
to the origin and ${\bf s}_j$ is the vector connecting 
the position of the $j$-th atomic nucleus of the molecule and 
the electron of the ion (as is viewed 
in the rest frame of the molecule).   

Using the first order perturbation theory in 
the interaction of the electron of the ion with 
the molecular field, described by the potentials (\ref{e4}), 
one can show that the cross section   
$\sigma_{fi}$ for the projectile-electron 
transitions occurring in collisions with the molecule 
is given by  
\begin{eqnarray}
\sigma_{fi} = 4 \sigma_{fi}^{(A)} %
\cos^2\left( \frac{ {\bf q}' \cdot {\bf l}_0 }{2} \right).   
\label{e12}  
\end{eqnarray} 
Here, $\sigma_{fi}^{(A)}$ is the cross section for 
the projectile-electron transitions occurring in collisions 
with the corresponding single atom, 
${\bf q}' = \left( {\bf q}_{\perp}, \frac{ \omega_{fi} }{ \gamma v } \right)$ 
is the momentum transferred to the projectile-ion 
(as viewed in the rest frame of the molecule) 
with $ \omega_{fi} $ being the transition frequency for the electron of the ion,   
and ${\bf l}_0 =(l_0, \vartheta_M, \varphi_M)$ is the vector connecting 
the positions of the atomic nuclei of the molecule 
in its rest frame. In what follows we shall count 
the polar orientation angle $\vartheta_M$ of the molecule 
from the direction of the projectile velocity ${\bf v}$. 
Besides, we set $\varphi_M = 0^0$. 

\begin{figure}[t] 
\vspace{-0.25cm}
\centering
\includegraphics[width=0.52\textwidth]{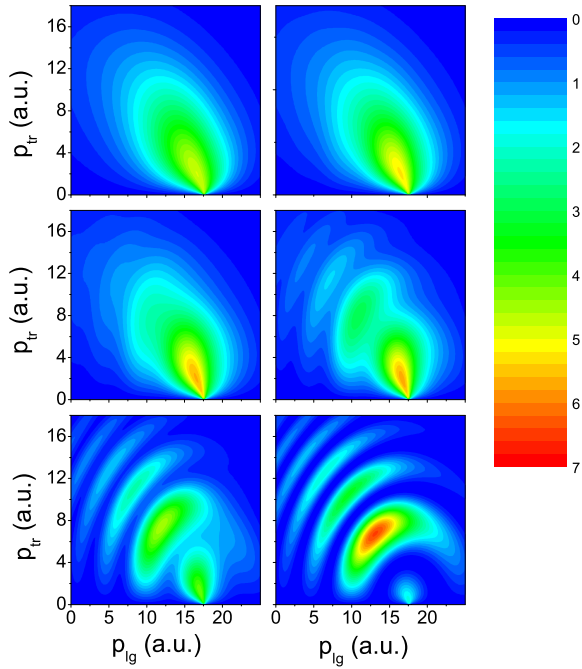}   
\vspace{-0.5cm}
\caption{ \footnotesize{ The spectra   
of electrons (in a.u.) emitted into the plane spanned 
by the molecular axis and projectile velocity 
from $7.8$ MeV/u Mg$^{11+}(1s)$ ions colliding 
with N$_2$ molecules.  
The spectra (from left to right, from top to bottom) 
correspond to $\vartheta_M  = 90^0$, $20^0$, $15^0$, 
$10^0$, $5^0$ and $0^0$.  
The angle $\vartheta_M$ is counted from the direction 
of the projectile velocity ${\bf v}$. } }  
\label{figure2}
\end{figure}  

In figure \ref{figure2} we present the electron 
loss cross section,  $d^3 \sigma/dp_{lg} dp_{tr} d\varphi_p$, 
differential in the longitudinal 
($p_{lg} = {\bf p} \cdot {\bf v}/v $) and 
transverse ($p_{tr} = \sqrt{ p^2 - p_{lg}^2 }$) 
momenta and the azimuthal emission angle $\varphi_p$ of the electrons 
emitted from $7.8$ MeV/u Mg$^{11+}(1s)$ projectiles   
in collisions with N$_2$ molecules. The cross section is 
obtained by integrating over the vector of the transverse momentum transfer 
${\bf q}_{\perp}$. In the figure this cross section is given  
in the target frame as a function of $p_{lg}$ and $p_{tr}$  
for the emission into the plane spanned by the molecular axis and 
projectile velocity (i.e. for $\varphi_p = 0^0 $).    
The molecular polar orientation angle is  
$\vartheta_M =90^0$, $20^0$ (the upper row, 
from left to right),  $15^0$, $10^0$ 
(the second row, from left to right)
and $5^0$, $0^0$ (the lower row, from left to right).  
At small $\vartheta_M$ the spectra exhibit very clear structures, 
which arise due to interference caused by the coherent interactions 
between the electron of the projectile and the two 
atomic centers of the molecule \cite{f1}.       
%

At an impact energy of $7.8$ MeV/u ($v=17.6$ a.u.)  
the typical momentum transfer 
to the electron of the ion, which is 
necessary for its removal out of the ion, 
is $ \sim 6$-$8$ a.u.. This magnitude is substantially larger 
than the typical momenta of the outer electrons of nitrogen.  
This means that within the screening target mode 
the projectile-electron transitions are governed 
mainly by the interaction between this electron and 
the target nuclei (partly screened by the $K$-shell electrons). 
Moreover, since the momentum transfers are large,  
the relative contribution of the collision mode, in which 
the target is excited,  
to the projectile-electron loss process 
is by about $Z_A=7$ times smaller than that due to   
the screening mode. Thus, the outer target electrons 
have a minor effect on the projectile-electron 
transitions and, therefore, the latter ones 
can indeed be regarded as occurring due to 
the interaction with two "slits", which are 
well localized and well separated from each other 
within the space occupied by the molecule. 

\begin{figure}[t] 
\vspace{-0.25cm}
\begin{center} 
\includegraphics[width=0.32\textwidth]{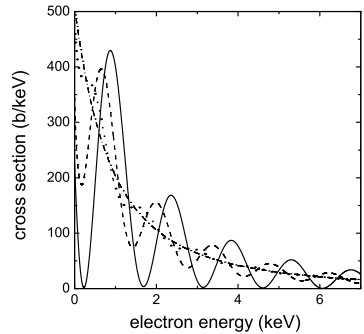} 
\end{center} 
\vspace{-0.75cm}  
\caption{ \footnotesize{ Energy spectrum  
of electrons (in a.u.) emitted under the zero azimuthal angle  
from $7.8$ MeV/u Mg$^{11+}$(1s) ions colliding with N$_2$. 
The spectrum is given in the projectile frame.  
Solid, dash and dot curves corresponds to collisions with the molecules 
oriented in the target frame under the polar angle 
$\vartheta_M = 0^0$, $5^0$ and $10^0$, respectively. 
For a comparison, dash-dot curve shows the spectrum 
in collisions with N atoms multiplied by $2$. } }  
\label{figure4} 
\end{figure}  

At small polar orientation angle of the molecule 
the spectrum displays clear ring-like structures. 
The center of the rings is 
located at the point ${\bf p}_C = (p_{tr}=0; p_{lg}= m_e v)$, 
where $m_e$ is the electron mass, implying 
that each ring is formed by electrons which 
in the rest frame of the projectile have close energies. 
Indeed, the origin of these structures 
can be traced back by considering the energy spectrum 
of the emitted electrons in the rest frame of the projectile. 
Such a spectrum is shown in figure \ref{figure4}.   
It is seen that in this frame the energy spectrum exhibits 
oscillations (especially pronounced at very small $\vartheta_M$) 
due to the alternation in the energy spectrum of the parts with constructive 
and destructive interferences. It is not difficult 
to convince oneself that the ring-like structures  
in the momentum spectrum originates namely from these oscillations.  

\begin{figure}[t] 
\vspace{-0.25cm} 
\centering 
\includegraphics[width=0.57\textwidth]{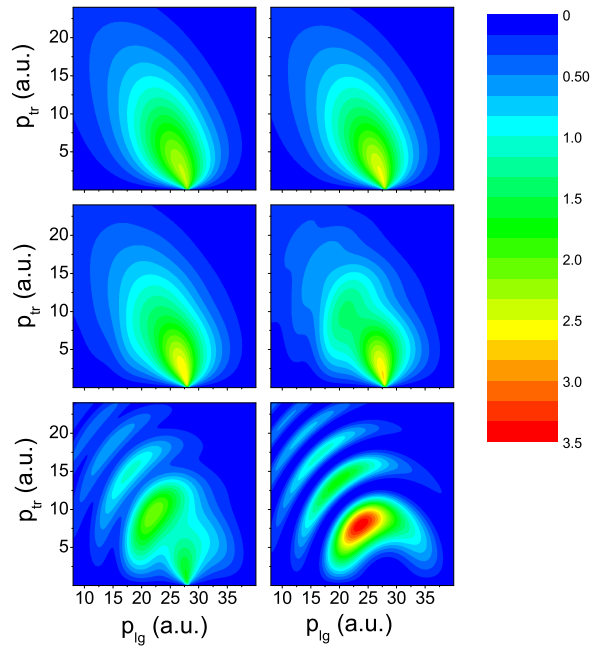}   
\vspace{-0.5cm}
\caption{ \footnotesize{ The same as in figure \ref{figure2}, 
but for $20$ MeV/u S$^{15+}(1s)$ ions colliding with N$_2$ molecules. } }  
\label{figure3}
\end{figure}  



For more information in figure \ref{figure3} 
we present the same cross section 
as in figure \ref{figure2}, but for 
the electron loss from $20$ MeV/u ($v=28.3$) S$^{15+}$(1s) 
projectiles colliding with N$_2$. 
Like in the previous case, the interference pattern  
in figure \ref{figure3} is caused by the coherent scattering  
of the electron of the projectile on the two "slits", which 
are very well localized in space and are distinctly separated 
from each other. 

Comparing the emission patterns in figures \ref{figure2} 
and  \ref{figure3} we see that the range of 
the molecular orientation angle $\vartheta_M$, 
at which the interference effects are clearly visible in the emission pattern, 
decreases when the charge $Z_p$ of the projectile nucleus increases. 
This can be easily understood if we recall that 
the size of the electron orbit in the initial state 
scales as $1/Z_p$. Therefore, a more tightly bound electron 
can interact simultaneously with both 
molecular centers only if the transverse size of the molecule 
$ l_{tr}=l_0 \sin \vartheta_M $ becomes smaller. 

As seen in figures 1--3, the most pronounced interference pattern in the 
emission spectrum arises at small orientation angles of the molecule. 
Therefore, in order to verify predicted effects in an experiment, it is 
very desirable to single out those loss events, which occur at 
small orientation angles, from the rest. This can be achieved 
by the determination of the molecular orientation 
ex post, which has been successfully applied in many experimental situations 
where molecular targets dissociated or Coulomb exploded after photo- and 
strong-field ionization or due to electron or ion impact induced 
ionization. 

In the collisions, considered above, 
by far a dominant contribution to the total electron emission 
is given by electrons ejected from the target. 
Therefore, an important question to address 
is whether in the momentum space there exists a substantial overlap 
between the electron emitted from the projectile 
and those ejected from  the target which would mask  
the above predicted interference effects.  
In order to answer it we have estimated the emission 
from the N$_2$ molecules. We found that, 
since $v > Z_p$ and $v \gg Z_A$, the overlap in the cases, 
considered in figures \ref{figure2}--\ref{figure3}, 
is small and the interference pattern 
in the electron emission from the projectile 
is not "damaged" by the electrons ejected from the target.  

In conclusion, we have considered interference 
effects in the electron emission accompanying 
energetic collisions of ionic projectiles with molecular targets.  
In contrast to all the previous studies of this subject, 
which were focused on interference in the electron emission 
from the target, we were searching for signatures of 
the interference effects in the electron emission from the projectile. 
We have shown that this emission  
may possess very clear interference structures  
which are caused by the coherent interactions  
between the electron of the projectile and  
the atomic centers of the molecule.  
Under certain conditions (which were discussed in detail above)     
this interaction is basically the one between 
the electron of the projectile and the nuclei of the atomic centers 
(partially screened by the inner shell atomic electrons). 
This means that the interference arises from the scattering 
of the projectile electron on atomic "slits", which are well 
localized in space and distinctly separated from each other,   
playing a role rather similar to that of the optical slits 
in the Young-type experiments with photons.  


Owing to recent advances in the experimental techniques 
it has become feasible to test the above theoretical predictions. 
In particular, this is planned to 
be done in forthcoming experiments at MPI-K (Heidelberg, Germany).


\begin{thebibliography}{99}
\bibitem{f1} Note that in the $90^0$-case 
the emission pattern is practically identical 
to that in collisions with two independent nitrogen atoms.
\end{thebibliography}
\end{document}